\documentclass[10pt,conference,final,twocolumn]{IEEEtran}

\usepackage{color}
\usepackage{bm}
\usepackage{bbm}
\usepackage{dsfont}
\usepackage{lipsum}

\usepackage[caption=1]{caption}
\usepackage{float}
\usepackage{stfloats}
\usepackage{eqparbox}
\usepackage{indentfirst}

\usepackage{fancyhdr}
\usepackage[left=0.7in,right=0.7in,top=0.77in,bottom=1.05in]{geometry}

\usepackage[style=ieee,maxnames=4,minnames=3,maxbibnames=3]{biblatex}

\usepackage{makeidx}

\usepackage{amsfonts}
\usepackage[cmex10]{amsmath}
\usepackage{amsthm}
\usepackage{amssymb}
\usepackage{mdwmath}
\usepackage{mathtools}
\usepackage{cases}

\theoremstyle{definition}

\theoremstyle{remark}

\theoremstyle{remark}

\theoremstyle{remark}

\theoremstyle{remark}

\usepackage{tabularx}
\usepackage{array}
\usepackage{booktabs}
\usepackage{makecell}
\usepackage{multirow}

\usepackage{comment}

\usepackage{algorithm}
\usepackage{algpseudocode}

\usepackage[pdftex]{graphicx}
\usepackage{epstopdf}
\usepackage{subcaption}
\graphicspath{ {./img/} }

\usepackage{url}

\usepackage{caption}
\begin{document}

\title{Geometry-Aided Channel Deduction: A Robust Channel Acquisition Framework Utilizing Coarse Scenario Prompt}
\author{Hongning~Ruan\textsuperscript{$\dagger$}, Zhaoyang~Zhang\textsuperscript{$\dagger\ddagger$}, Zirui~Chen\textsuperscript{$\dagger$}, Ziqing~Xing\textsuperscript{$\dagger\ddagger$}, and Zhaohui~Yang\textsuperscript{$\dagger$}\\
    \IEEEauthorblockA{
        \textsuperscript{$\dagger$}College of Information Science and Electronic Engineering, Zhejiang University, Hangzhou 310027, China\\
        \textsuperscript{$\dagger$}Zhejiang Provincial Laboratory of Multi-Modal Communication Networks\\and Intelligent Information Processing, Hangzhou 310027, China\\
        \textsuperscript{$\ddagger$}Institute of Fundamental and Transdisciplinary Research, Zhejiang University, Hangzhou 310058, China\\
        E-mails: \{rhohenning, ning\_ming, ziruichen, ziqing\_xing, yang\_zhaohui\}@zju.edu.cn
    }
}

\date{}
\maketitle

\begin{abstract}
Channel state information (CSI) is critical for multi-input multi-output (MIMO) orthogonal frequency division multiplexing (OFDM) system. Pilot-based channel estimation methods suffer from high pilot overhead and low channel acquisition quality, while pilot-free approaches typically impose impractical demands on positional or environmental information precision. This paper proposes geometry-aided channel deduction (GCD), which leverages readily available geometric information to assist channel acquisition. 
The environmental map and base station position together constitute the scenario geometry, which can provide geometric channel features through ray tracing. To obtain the complete channel, the user first retrieves approximate geometric features by performing neighborhood searching within a pre-extracted geometric feature set, and then converts them into pseudo channels through a priori designed feature alignment. These pseudo channels serve as contextual prompt, providing supplementary channel features beyond those derived from pilot-based estimate. Finally, a neural network fuses these pseudo channels with partial estimate to generate the complete channel. 
Comprehensive experiments validate the superiority of our method, which achieves the leading accuracy in channel acquisition under sparse pilot conditions, demonstrates strong generalization capabilities in new scenarios and dynamic environments, and exhibits robust resilience against user position errors and non-ideal environmental information.
\end{abstract}


\section{Introduction}

In wireless communication systems, precise channel acquisition is critical because channel state information (CSI) directly impacts system performance. However, multiple-input multiple-output (MIMO) and orthogonal frequency division multiplexing (OFDM) technologies substantially increase the volume of channel data. Reliably acquiring high-dimensional CSI in complex urban environments with limited pilot overhead poses particular challenges.

Deep learning (DL) excels at uncovering latent features and representing high-dimensional data and has been applied to the field of channel estimation. For instance, \cite{chen2024cmixer} proposed the complex-domain multi-layer perceptron (MLP)-Mixer (CMixer) that employs physics-inspired design to map pilot-based partial estimates to complete CSI, outperforming previous approaches based on MLP or convolutional neural network (CNN). However, relying solely on limited pilot resources typically fails to provide sufficient channel features when the multi-path characteristics are highly complex.

Some studies focus on achieving pilot-free channel acquisition by learning deterministic mappings that convert other available information into complete channel data. For instance, the current channel can be predicted from past channels, known as channel prediction \cite{jiang2022prediction}. However, the randomness in user's movement is typically unpredictable, and the prediction errors are accumulated over time during auto-regressive inference, leading to severe performance degradation. Some approaches utilize user positions to obtain channels \cite{xiao2022cgrbfnet,chatelier2025loc2ch}, but these methods not only demand stringent positioning accuracy but also lack generalization capability in new scenarios. Other studies focus on reconstructing the physical environment \cite{hoydis2024learning}, as environmental information directly determines the wireless channel. However, these approaches still rely on high-precision user positioning and environmental geometry. They also require calibration of environmental materials, necessitating data collection and computation for each scenario. Furthermore, all these methods neglect the diversity of antenna radiation characteristics, which is influenced by numerous factors such as hardware configurations and user device orientations.

Recent studies enhance channel estimation by combining pilot resources with auxiliary information. On one hand, the supplementary channel features provided by the auxiliary information can significantly reduce pilot overhead; on the other hand, by introducing tiny pilot resources, these methods relax the accuracy requirements for auxiliary information, as some small-scale or unpredictable features can be calibrated by real-time estimates from pilots. For instance, \cite{chen2025scd} proposed position-aided CMixer (PCMixer) that fuses position and channel information to obtain the complete channel, thereby tolerating positioning errors. \cite{chen2025cd} proposed a channel deduction framework that combines estimation with prediction, leveraging past channels to provide large-scale features. To obtain more precise and context-aware features, recent studies incorporate scenario information such as historical channel datasets \cite{chen2025scd}, digital twin channel \cite{cai2025ecbp2wcp}, and real-time captured images \cite{shi2025weicp}. However, these methods typically require costly data collection, transmission, or storage, which is particularly pronounced when deployed in new scenarios.

Existing 3D reconstruction techniques facilitate the acquisition of static environmental maps. This has motivated us to leverage readily available geometric information -- the environmental map and transceiver positions -- to assist channel estimation. Our approach not only significantly reduces pilot overhead but also introduces scenario-related context, enabling our method to achieve strong generalization in new scenarios \cite{chen2025al}. Furthermore, the low cost and relaxed accuracy requirements for acquiring geometric information enhance the practical feasibility of our approach in real-world systems. The main contributions of this paper are as follows:

\begin{itemize}
\item By theoretically analyzing the channel model, we identify readily obtainable channel features and consequently propose the geometry-aided channel deduction (GCD) framework that fuses geometric features with pilot-based partial estimate to derive the complete channel.
\item We present a detailed implementation of the proposed framework, which employs ray tracing and neighborhood sampling techniques to extract geometric features, and introduces a feature alignment mechanism to facilitate the neural network's fusion of channel information.
\item We conduct comprehensive experiments to evaluate our method, revealing its exceptional channel acquisition accuracy, strong robustness against non-ideal geometric information, and remarkable cross-scenario generalization capability.
\end{itemize}

\section{System Model}

\subsection{Channel Model}

According to electromagnetic theory, the single-frequency channel response between a transmitter (TX) and a receiver (RX) can be expressed as:
\begin{equation}
h=\sum_{p=1}^P\underbrace{\frac{\lambda}{4\pi d_p}\mathbf{C}_{\rm R}(\widehat{\bm{k}}_{{\rm R},p})^{\mathsf{H}}\bm{\Xi}_p\mathbf{C}_{\rm T}(\widehat{\bm{k}}_{{\rm T},p})}_{\triangleq\alpha_p}e^{-j2\pi f\tau_p},
\end{equation}
where $f$ is the carrier frequency, $\lambda=c/f$ is the wavelength, and $c$ denotes the speed of light.

The channel from TX to RX comprises $P$ propagation paths, each possessing several attributes: path length $d_p$, departure direction $\widehat{\bm{k}}_{{\rm T},p}$, arrival direction $\widehat{\bm{k}}_{{\rm R},p}$, and $3\times 3$ scattering matrix $\bm{\Xi}_p$ describing the effect of electromagnetic interaction with the scattering environment. The propagation delay of each path is defined as $\tau_p=d_p/c$. Additionally, the $3\times 1$ vectors $\mathbf{C}_{\rm T}(\widehat{\bm{k}}_{{\rm T},p})$ and $\mathbf{C}_{\rm R}(\widehat{\bm{k}}_{{\rm R},p})$ represent the direction-dependent antenna patterns of the TX and RX, which describes the antennas' amplitude gains, initial phases, and polarization directions. For simplicity, the overall amplitude attenuation of each path is denoted as $\alpha_p$.

We consider a MIMO-OFDM system, where a base station (BS) equipped with $N_{\rm t}$ antennas serves single-antenna users via $N_{\rm c}$ subcarriers. The downlink channel from the BS to a singe user is represented by $\mathbf{H}\in\mathbb{C}^{N_{\rm t}\times N_{\rm c}}$, defined as
\begin{equation}\label{eq::mimo_ofdm_channel}
\mathbf{H}[n_{\rm t},n_{\rm c}]=\sum_{p=1}^P\alpha_pe^{-j2\pi f\tau_p}e^{-j2\pi n_{\rm c}\Delta_f\tau_p}e^{jk\bm{d}_{n_{\rm t}}^{\mathsf{T}}\widehat{\bm{k}}_{{\rm T},p}}e^{j2\pi\nu_pt},
\end{equation}
where $n_{\rm c}\Delta_f$ is the frequency offset of the $n_{\rm c}$-th subcarrier, $\Delta_f$ is the subcarrier spacing, $k=2\pi/\lambda$ is the wavenumber, $\bm{d}_{n_{\rm t}}$ is the position of the $n_{\rm t}$-th BS antenna relative to the reference antenna, and $\nu_p=\bm{v}^{\mathsf{T}}\widehat{\bm{k}}_{{\rm R},p}/\lambda$ is the Doppler frequency of each path contributed by the user velocity $\bm{v}$.

\subsection{Obtainable Channel Features}

The path lengths $d_p$ and the departure and arrival directions $\widehat{\bm{k}}_{{\rm T},p}$, $\widehat{\bm{k}}_{{\rm R},p}$ constitute the geometric features of a multi-path channel. These parameters can be obtained using ray tracing as long as the environmental map and the positions of BS and users are available. In practical applications, the BS can acquire the environmental map through 3D reconstruction techniques or geographic information system (GIS), while the BS and users can obtain their respective positions via global navigation satellite system (GNSS). For privacy reasons, the BS and users may be reluctant to disclose their positions externally. Furthermore, the acquired geometric information may exhibit errors exceeding the wavelength scale, e.g., geometric accuracy limited to the meter level. Such errors render the high-frequency phase term $e^{-j2\pi f\tau_p}$ unpredictable.

The scattering matrices $\bm{\Xi}_p$ depend on the electromagnetic properties of environmental materials. The antenna patterns $\mathbf{C}_{\rm T}(\widehat{\bm{k}}_{{\rm T},p})$ and $\mathbf{C}_{\rm R}(\widehat{\bm{k}}_{{\rm R},p})$ depend on hardware configurations and antenna orientations. The Doppler frequencies $\nu_p$ depend on the user mobility. We assume these factors are difficult to obtain with precision.

The center frequency $f$, subcarrier spacing $\Delta_f$, and relative positions of BS antennas $\bm{d}_{n_{\rm t}}$ are all pre-defined system configurations. These parameters can be obtained with precision.

Based on the above discussion, we decompose the channel response of each path into two parts, rewriting \eqref{eq::mimo_ofdm_channel} as follows:
\begin{align}\label{eq::mimo_ofdm_channel_2parts}
\mathbf{H}[n_{\rm t},n_{\rm c}]=\sum_{p=1}^P
&\ \underbrace{\frac{\lambda}{4\pi d_p}e^{-j2\pi n_{\rm c}\Delta_f\tau_p}e^{jk\bm{d}_{n_{\rm t}}^{\mathsf{T}}\widehat{\bm{k}}_{{\rm T},p}}}_{\text{obtainable}}\notag\\
&\ \cdot\underbrace{\mathbf{C}_{\rm R}(\widehat{\bm{k}}_{{\rm R},p})^{\mathsf{H}}\bm{\Xi}_p\mathbf{C}_{\rm T}(\widehat{\bm{k}}_{{\rm T},p})e^{-j2\pi f\tau_p}e^{j2\pi\nu_pt}}_{\text{unobtainable}}.
\end{align}
The first part is obtainable as it consists of geometric features and fixed system configurations. Conversely, the second part is typically unobtainable.

\begin{figure}[t]
\centering
\includegraphics[width=\columnwidth]{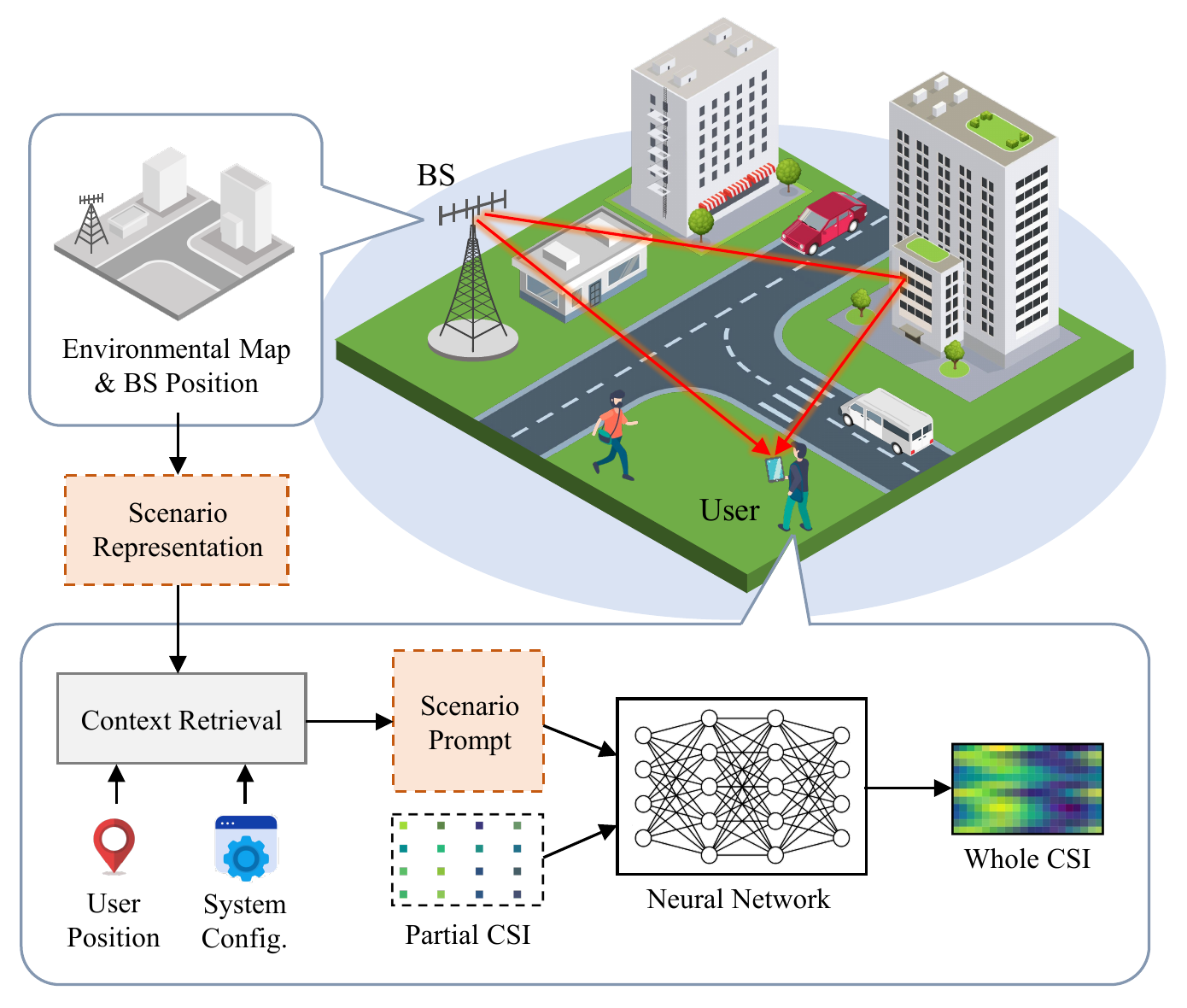}
\caption{Overview of the proposed channel acquisition framework.}
\label{fig::overview}
\vspace{-0.1cm}
\end{figure}

\begin{figure*}[t]
\centering
\includegraphics[width=2\columnwidth]{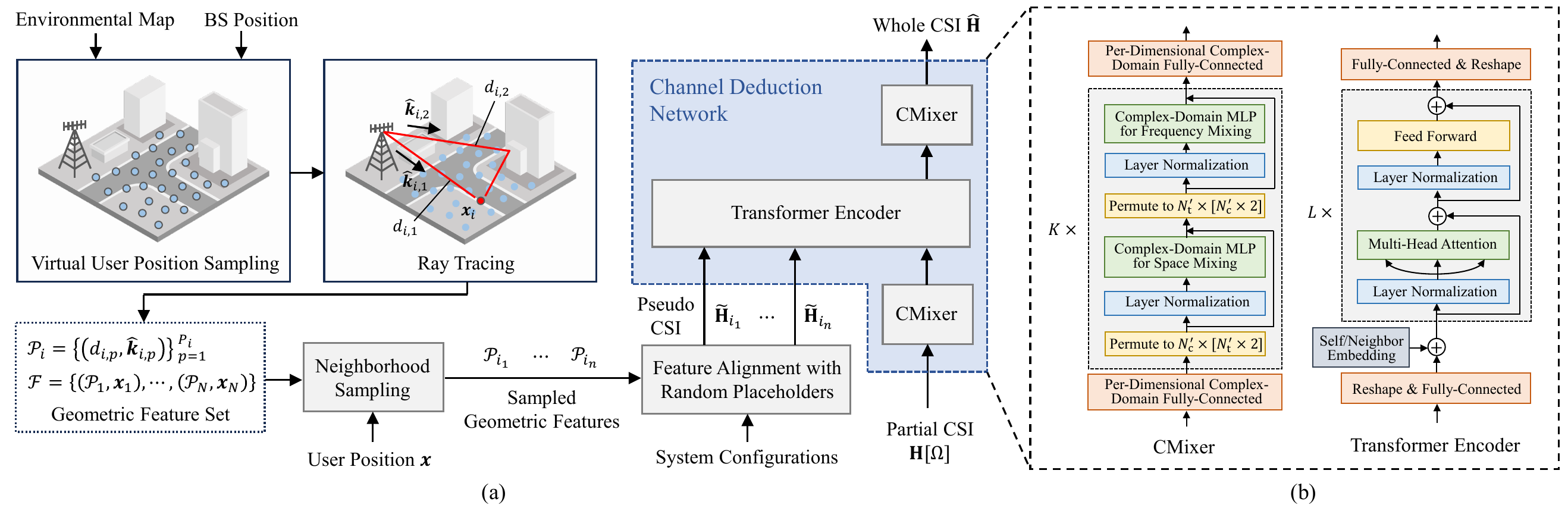}
\caption{(a) Detailed diagram of our approach. (b) CMixer and Transformer encoder used in the channel deduction network.}
\label{fig::framework}
\vspace{-0.1cm}
\end{figure*}

\section{Proposed Framework}

\subsection{Overview}
To leverage available channel features and enhance channel estimation efficiency, we propose a novel channel acquisition framework named geometry-aided channel deduction (GCD). Fig. \ref{fig::overview} presents an overview of this framework. The BS possesses its own position and a coarse environmental map, which collectively form the scenario representation. The user retrieves geometric channel features from the scenario representation based on its own position $\bm{x}$ and converts them into the CSI modality using known system configurations, thereby generating a scenario prompt that provides contextual information. Meanwhile, the partial channel $\mathbf{H}[\Omega]\in\mathbb{C}^{N_{\rm t}^0\times N_{\rm c}^0}$ is estimated through known pilots, where $\Omega=\Omega_{\rm t}\times\Omega_{\rm c}$ is a subset of antenna-subcarrier indices, $\times$ denotes the Cartesian product, and $\Omega_{\rm t}\subseteq\{0,1,\cdots,N_{\rm t}-1\}$, $|\Omega_{\rm t}|=N_{\rm t}^0<N_{\rm t}$, $\Omega_{\rm c}\subseteq\{0,1,\cdots,N_{\rm c}-1\}$, $|\Omega_{\rm c}|=N_{\rm c}^0<N_{\rm c}$. Subsequently, a neural network fuses the partial channel with the scenario prompt to generate the complete channel $\widehat{\mathbf{H}}$.

The aforementioned inference process resembles retrieval-augmented generation (RAG) for large language models \cite{gao2023retrieval}. The scenario representation functions similarly to a knowledge base containing domain-specific information. Contextual prompts are constructed from the knowledge base according to user queries and augment the original inputs to generate more accurate and comprehensive outputs. However, unlike human language that can be universally converted into vectors via word embedding, in our problems, the inputs and outputs are wireless channel responses while the external knowledge consists of only geometric path parameters. They belong to different modalities and share an intrinsic structural correlation dictated by the channel model. Therefore, an efficient and physics-constrained approach is required to align the geometric features to the complete channel representation \cite{chen2025towards}.

The following subsections detail the implementations of our framework, which is illustrated in Fig. \ref{fig::framework}.

\subsection{Geometric Feature Retrieval}

Given the environmental map and transceiver positions, geometric channel features can be obtained through ray tracing. However, directly applying ray tracing presents several drawbacks. First, ray tracing requires positional information from both the BS and the user, raising privacy concerns -- at least one of them must disclose its position to the other. Furthermore, the user position is constantly changing, meaning ray tracing must be repeatedly performed based on real-time user position. This consumes substantial computational resources and results in significant inference latency.

Considering these issues, we enable the BS to pre-extract geometric features offline without requiring actual user positions. It first samples $N$ virtual user positions $\bm{x}_1,\cdots,\bm{x}_N$ in the environmental map, which are supposed to approximately cover the activity area of potential users. Ray tracing is then performed using these positions, obtaining the geometric feature set $\mathcal{F}=\{(\mathcal{P}_1,\bm{x}_1),\cdots,(\mathcal{P}_N,\bm{x}_N)\}$, where $\mathcal{P}_i=\{(d_{i,p},\widehat{\bm{k}}_{i,p})\}_{p=1}^{P_i}$ contains the path parameters at position $\bm{x}_i$, $P_i$ denotes the number of paths from the BS to $\bm{x}_i$, and $d_{i,p}$, $\widehat{\bm{k}}_{i,p}$ denote the length and departure direction of each path, respectively. $\mathcal{F}$ can be regarded as a compact representation of the scenario, which can be provided to users without revealing the actual BS position.

Given the geometric feature set $\mathcal{F}$, the user searches for the nearest neighbors among the virtual user positions $\bm{x}_1,\cdots,\bm{x}_N$ within $\mathcal{F}$ based on its own position $\bm{x}$. We denote the $n$ nearest virtual user positions as $\bm{x}_{i_1},\cdots,\bm{x}_{i_n}$ and their corresponding geometric features as $\mathcal{P}_{i_1},\cdots,\mathcal{P}_{i_n}$. Given the high similarity of scattering environment within the spatial neighborhood, the geometric features at position $\bm{x}$ can be well approximated by those at its neighbors.

\subsection{Feature Alignment}

As previously mentioned, the available geometric features and partial estimate are fused to obtain the complete channel, where the partial channel compensates for the remaining inaccessible features. To ensure more learnable intra-modal fusion, we transform the geometric path parameters $\mathcal{P}_i$ ($i=i_1,\cdots,i_n$) into the form of complete channels. Specifically, we introduce a random placeholder $z_{i,p}\sim\mathcal{CN}(0,\sigma_{\rm z}^2)$ for each path to replace the unobtainable part in \eqref{eq::mimo_ofdm_channel_2parts}, where $\sigma_{\rm z}$ is set to 0.5 in our experiment. Consequently, geometric features $\mathcal{P}_i$ at virtual user position $\bm{x}_i$ can be used to construct a pseudo channel $\widetilde{\mathbf{H}}_i\in\mathbb{C}^{N_{\rm t}\times N_{\rm c}}$ as follows:
\begin{equation}
\widetilde{\mathbf{H}}_i[n_{\rm t},n_{\rm c}]=\sum_{p=1}^{P_i}\frac{\lambda}{4\pi d_{i,p}}z_{i,p}e^{-j2\pi n_{\rm c}\Delta_f\tau_{i,p}}e^{jk\bm{d}_{n_{\rm t}}^{\mathsf{T}}\widehat{\bm{k}}_{i,p}}.
\end{equation}
The pseudo channels $\widetilde{\mathbf{H}}_{i_1},\cdots,\widetilde{\mathbf{H}}_{i_n}$ derived from neighboring samples provide approximate geometric features of the target channel and can therefore serve as the contextual prompt.

The random placeholders introduced here serve multiple purposes. First, they supplement unknown information within the channels, aligning known geometric features to the complete channel representation. This enables the subsequent neural network to fuse different channel information within the same feature space. Second, by introducing independent random placeholders for different pseudo channels, the neural network avoids overfitting to the useless information represented by these placeholders, thereby focusing on useful geometric features common to all pseudo channels.

\subsection{Channel Deduction Network}

We employ the attention-based channel deduction network (ACDNet) proposed in \cite{chen2025cd} to fuse the pseudo channels with the partial channel. As shown in Fig. \ref{fig::framework}(a), the partial channel $\mathbf{H}[\Omega]$ is input into a $K$-layer CMixer to produce an output with the shape of the full channel, which is subsequently fused with the pseudo channels $\widetilde{\mathbf{H}}_{i_1},\cdots,\widetilde{\mathbf{H}}_{i_n}$ through an $L$-layer Transformer encoder. The resulting output is then processed by another $K$-layer CMixer to generate the full channel $\widehat{\mathbf{H}}$. The network structures of CMixer and Transformer encoder are illustrated in Fig. \ref{fig::framework}(b). In our experiment, we set $K=3$ and $L=6$, with the hidden size of Transformer encoder set to $N_{\rm t}N_{\rm c}/2$.

To mitigate the impact of the absolute magnitude of channel responses, we normalize the channel data before feeding it into the network. Specifically, we first estimate the power using known partial channel, denoted as $P_{\mathbf{H}}=\|\mathbf{H}[\Omega]\|_{\rm F}^2/(N_{\rm t}^0N_{\rm c}^0)$, and then perform the normalization operation: $\mathbf{H}\gets\mathbf{H}/\sqrt{P_{\mathbf{H}}}$, $\mathbf{H}[\Omega]\gets\mathbf{H}[\Omega]/\sqrt{P_{\mathbf{H}}}$, $\widetilde{\mathbf{H}}_i\gets\widetilde{\mathbf{H}}_i/\sqrt{P_{\mathbf{H}}}$ ($i=i_1,\cdots,i_n$). The network is trained using the Adam optimizer, with the loss function defined as $\mathcal{L}=\|\mathbf{H}-\widehat{\mathbf{H}}\|_{\rm F}^2/(N_{\rm t}N_{\rm c})$. During inference, the acquired channel is restored via $\widehat{\mathbf{H}}\gets\sqrt{P_{\mathbf{H}}}\widehat{\mathbf{H}}$.

\begin{figure}[t]
\centering
\includegraphics[width=0.9\columnwidth]{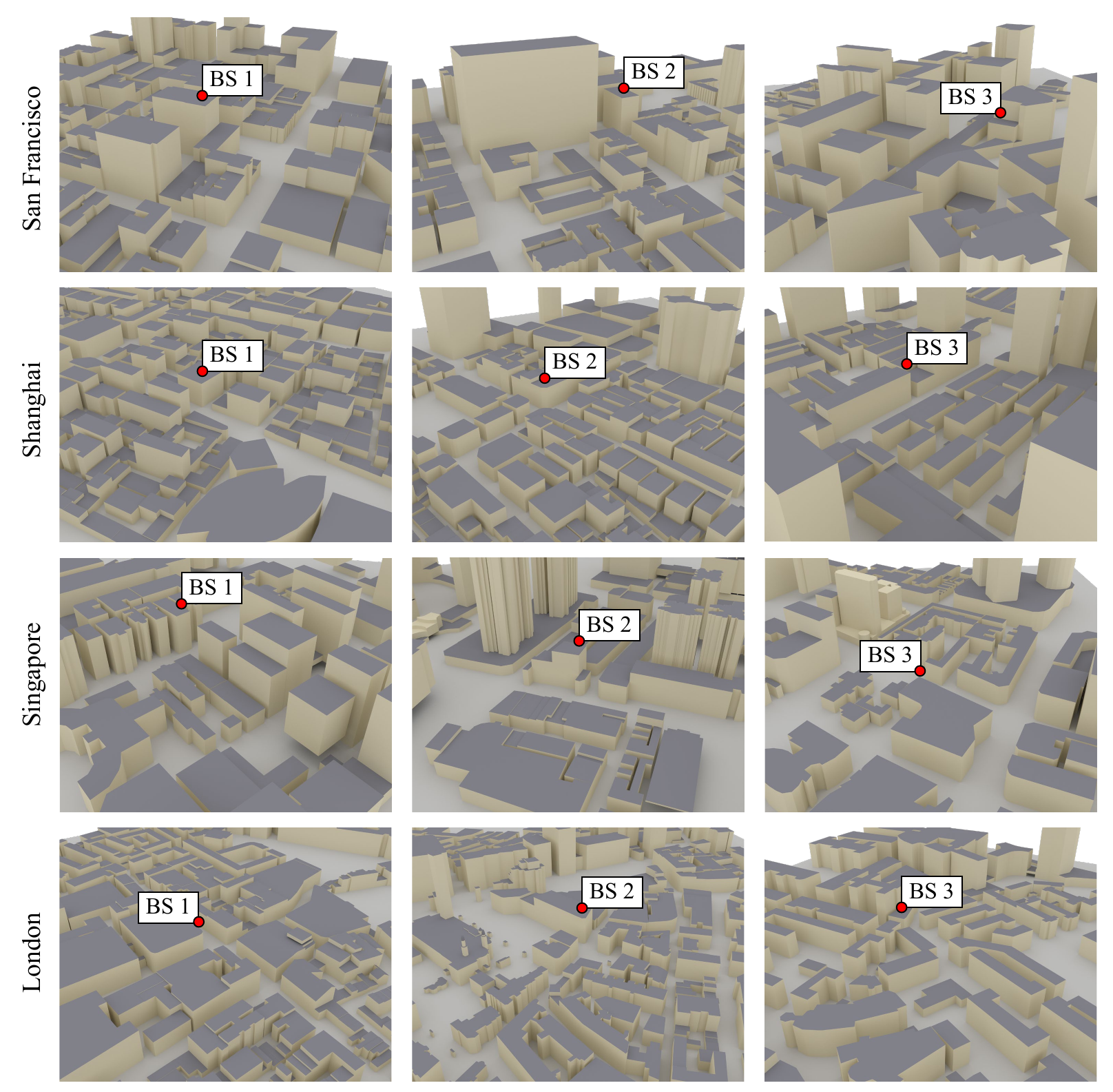}
\caption{Environmental maps of the 12 scenarios.}
\label{fig::scenes}
\vspace{-0.1cm}
\end{figure}

\begin{table}[t]
\centering
\caption{Settings of system parameters.}
\label{tab::setting}
\begin{tabular}{ll}
\toprule
\textbf{Parameter} & \textbf{Value} \\
\midrule
Center frequency $f$ & 5GHz \\
Bandwidth $B=N_{\rm c}\Delta_f$ & 40MHz \\
Number of subcarriers $N_{\rm c}$ & 256 \\
Number of BS antennas $N_{\rm t}$ & 16 \\
Form of BS antenna array & Uniform linear array (ULA) \\
Partial channel size $N_{\rm t}^0\times N_{\rm c}^0$ & $4\times 16$ \\
Subset $\Omega=\Omega_{\rm t}\times\Omega_{\rm c}$ & $\{0,4,8,12\}\times\{0,16,32,\cdots,240\}$ \\
\bottomrule
\end{tabular}
\end{table}

\section{Simulation Results}

\subsection{Experimental Settings}

We collect the environmental maps from OpenStreetMap for four global cities: San Francisco, Shanghai, Singapore, and London. For each city, we manually set three BS positions, constructing 12 scenarios, as shown in Fig. \ref{fig::scenes}. Densely distributed buildings create numerous non-line-of-sight (NLoS) areas, resulting in highly complex multi-path characteristics that pose challenges for channel acquisition. The environmental maps are imported into Sionna to generate channel data and geometric features. In each scenario, the users are randomly distributed within a square area centered on the BS, measuring $400\text{m}\times 400\text{m}$. User heights range from $1\text{m}$ to $2\text{m}$. Virtual user positions are sampled on regular grids with $1\text{m}\times 1\text{m}$ grid size and a height of $1.5\text{m}$. All users are equipped with randomly oriented dipole antennas to simulate the diverse antenna radiation characteristics. Other system parameters are detailed in Table \ref{tab::setting}. Notably, the known pilots take up only $(N_{\rm t}^0N_{\rm c}^0)/(N_{\rm t}N_{\rm c})=1/64$ of total channel resources, making it challenging to provide sufficient information for traditional channel estimation.

We select CMixer \cite{chen2024cmixer} and PCMixer \cite{chen2025scd} as two baselines. CMixer consists of 8 layers, generating the complete channel using only pilot-based estimates. PCMixer first processes the user position and partial channel separately through two 6-layer ResMLPs, followed by $1\times 1$ convolutional fusion, and finally generates the full channel via an 8-layer CMixer. We set the hidden size of the two ResMLPs in PCMixer to $N_{\rm t}N_{\rm c}/2$. All networks are trained for 1000 epochs with a batch size of 500. The learning rate is initially set to $10^{-4}$ and decayed by a factor of 0.8 every 100 epochs. During GCD training, the number of pseudo channels $n$ is randomly selected from 0 to 16. During testing, $n$ is set to 16 by default.

We evaluate the channel acquisition quality for each data sample using normalized mean squared error (NMSE), which is defined as $\text{NMSE}=\|\mathbf{H}-\widehat{\mathbf{H}}\|_{\rm F}^2/\|\mathbf{H}\|_{\rm F}^2$.

\begin{figure*}[t]
\centering
\includegraphics[width=2\columnwidth]{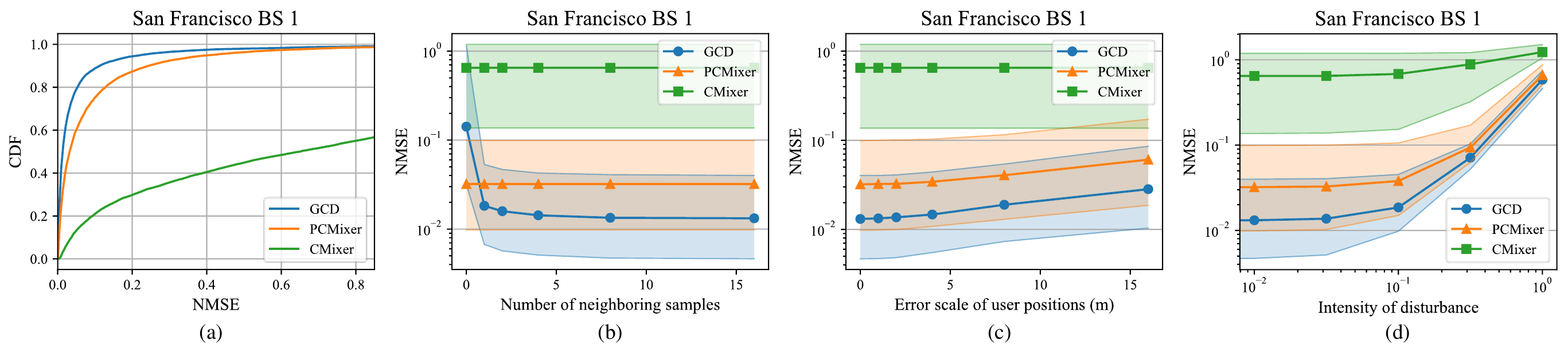}
\caption{Experimental results of single-scenario learning. (a) Cumulative probability distribution of NMSE. (b) NMSE under different numbers of neighboring samples. (c) NMSE under different positioning errors. (d) NMSE under different noise levels of partial channel. The median NMSE and quartiles are presented. }
\label{fig::results_sf1}
\vspace{-0.1cm}
\end{figure*}

\subsection{Single-Scenario Learning}

We first train and evaluate all schemes in a single scenario -- San Francisco BS 1. The training, validation, and testing datasets comprise 40k, 10k, and 10k data samples, respectively. Fig. \ref{fig::results_sf1}(a) displays the cumulative distribution of NMSE. Due to the complex multi-path characteristics and limited pilot resources, CMixer struggles to perform effectively. In contrast, both PCMixer and GCD incorporate additional information, achieving superior accuracy, with GCD demonstrating particularly outstanding performance. Fig. \ref{fig::results_sf1}(b) shows GCD's performance under various numbers of neighboring samples $n$. As $n$ increases, GCD's performance improves first and then saturates. Even with only one neighboring sample, GCD still achieves superior performance over other schemes.

In the following, we conduct several experiments to assess the robustness of our method. We first introduce positioning errors by generating inaccurate user positions as $\widehat{\bm{x}}=\bm{x}+\Delta\bm{x}$, where $\Delta\bm{x}=[c,d]$, $c,d$ are uniformly sampled within the interval $[-l,l]$, with $l$ being the error scale\footnote{Here, we represent the user position as a 2D vector, omitting height information. Since user heights typically fall within in a narrow range (e.g., $[1\text{m},2\text{m}]$), we simply assign all virtual user positions the same height ($1.5\text{m}$). Consequently, real user height values are unnecessary here as they do not affect the neighborhood searching result.}. The results are shown in Fig. \ref{fig::results_sf1}(c). Both GCD and PCMixer demonstrate strong robustness against positioning errors, with GCD outperforming other methods under all conditions. Next, we apply disturbance to known partial channels by generating the noisy estimate as $\mathbf{H}'[\Omega]=\mathbf{H}[\Omega]\odot\mathbf{D}$, where $\odot$ denotes the Hadamard product, $\mathbf{D}$ is the disturbance matrix with elements independently sampled from $\mathcal{N}(1,\sigma_{\mathbf{D}}^2)$, and $\sigma_{\mathbf{D}}$ is the disturbance intensity. As shown in Fig. \ref{fig::results_sf1}(d), GCD consistently outperforms other methods across various noise levels.

\begin{figure}[t]
\centering
\includegraphics[width=0.85\columnwidth]{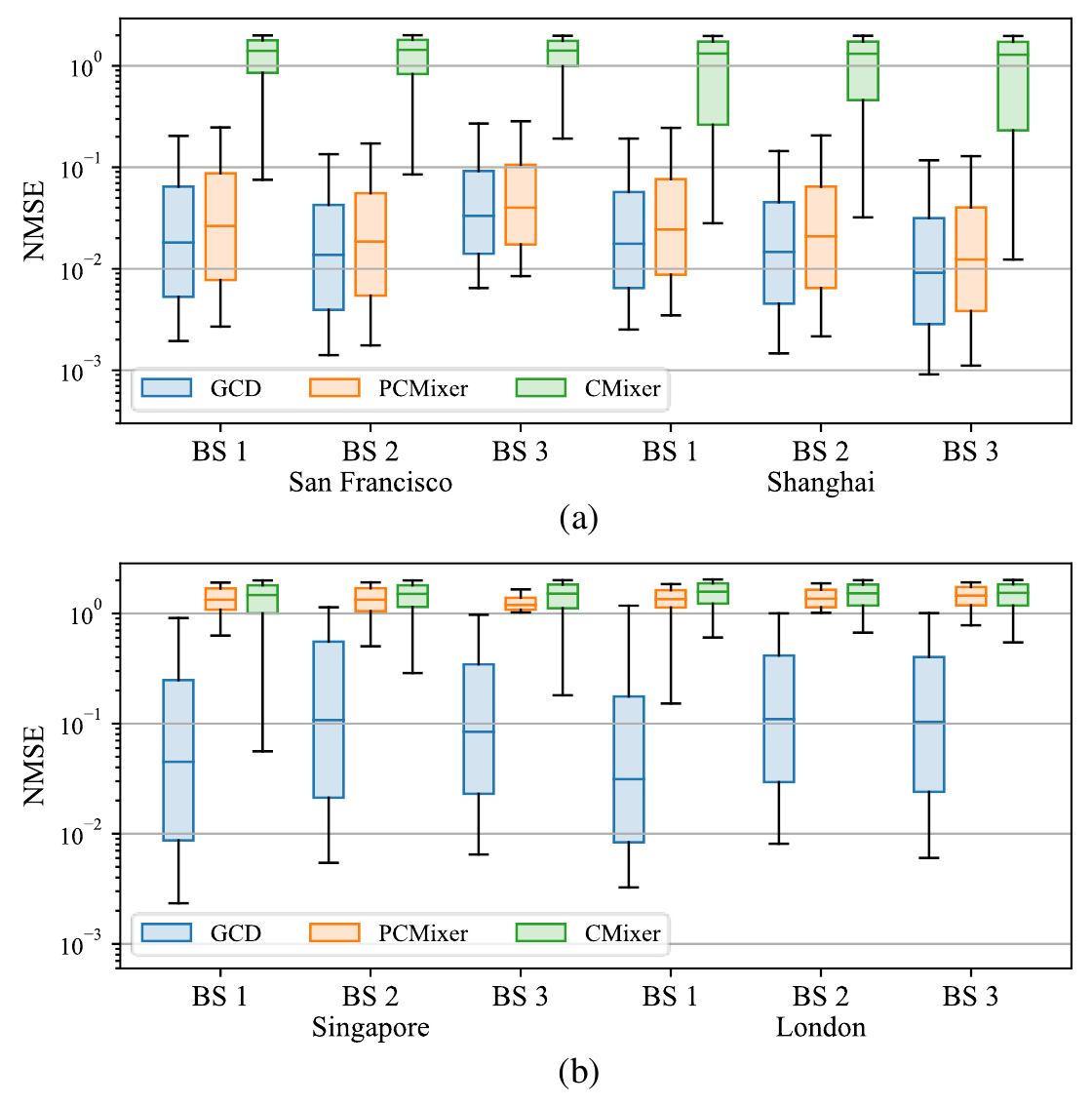}
\caption{Experimental results of multi-scenario learning. (a) NMSE performance in scenarios involved in training. (b) NMSE performance in unseen scenarios without additional training.
The whiskers extend from the 10th percentile to the 90th percentile, and flier points are omitted.
}
\label{fig::results_all}
\vspace{-0.1cm}
\end{figure}

\subsection{Multi-Scenario Learning}

In this subsection, we train each scheme using training data from the first 6 scenarios jointly and evaluate it across all 12 scenarios. The training, validation, and testing datasets for each scenario contain 20k, 10k, and 10k data samples, respectively. Fig. \ref{fig::results_all}(a) shows the NMSE performance across the first 6 scenarios involved in the training process. CMixer fails in multi-scenario learning due to highly diverse channel characteristics and limited pilot resources, while both PCMixer and GCD achieve multi-scenario collaborative learning, with GCD consistently outperforming PCMixer. Fig. \ref{fig::results_all}(b) shows the NMSE performance across the remaining 6 scenarios, where all schemes are evaluated directly without additional training or finetuning. PCMixer and CMixer completely fails in these unseen scenarios. In contrast, GCD demonstrates remarkable generalization capability. Despite performance degradation due to substantial differences in data distribution compared to previous scenarios, GCD achieves satisfactory NMSE performance on most testing samples in the new scenarios.

\begin{figure}[t]
\centering
\includegraphics[width=0.95\columnwidth]{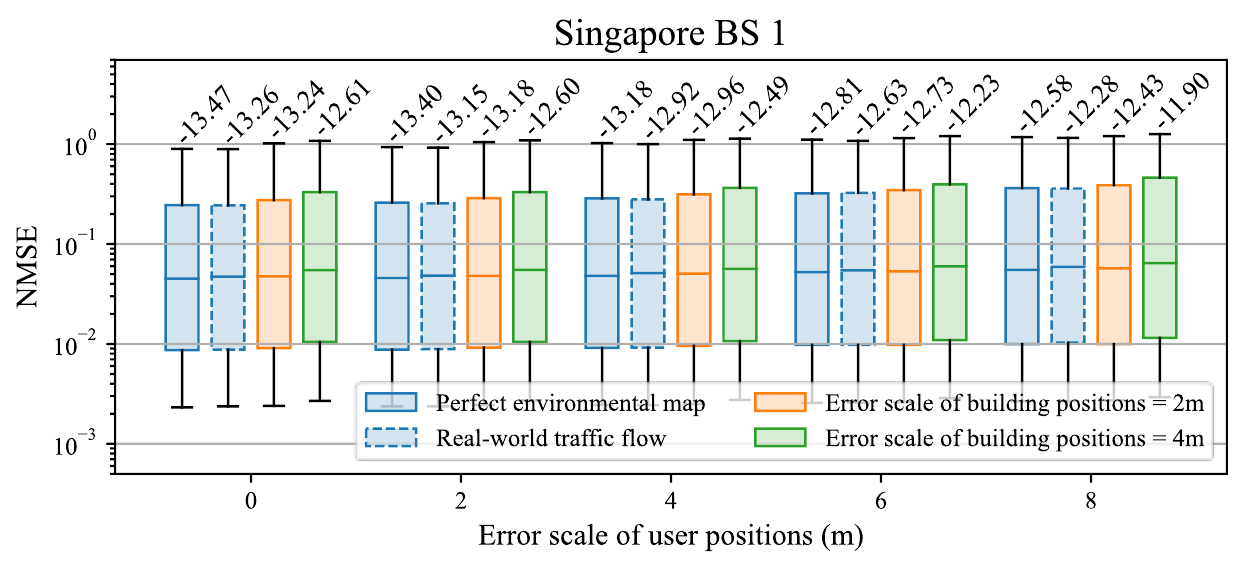}
\caption{NMSE performance of GCD using non-ideal geometric information. The median NMSE in dB is provided above the whisker.}
\label{fig::results_biased}
\vspace{-0.1cm}
\end{figure}

Next, we evaluate GCD's robustness against non-ideal environmental geometry. We consider the following two cases:
\begin{itemize}
\item \textbf{Real-world traffic flow:} There are moving vehicles that affect the real-time channel, but the environmental map captures only static buildings. In our experiment, we manually place several vehicles in the environmental map and then regenerate channel data.
\item \textbf{Building position errors:} The building positions in the environmental map are biased. Similar to generating inaccurate user positions, we independently generate inaccurate positions for each building in the environmental map and then regenerate geometric features.
\end{itemize}
Fig. \ref{fig::results_biased} illustrates GCD's performance in one of the unseen scenarios -- Singapore BS 1. GCD demonstrates strong robustness across all types of geometric information errors.

\begin{figure}[t]
\centering
\includegraphics[width=\columnwidth]{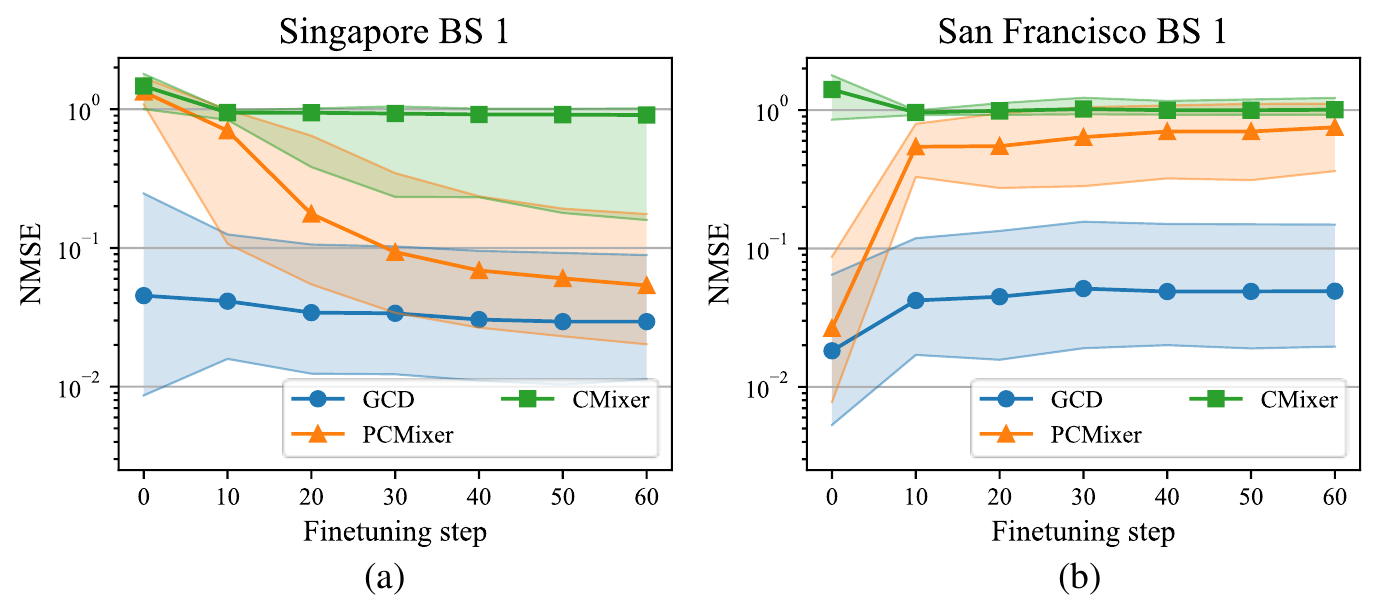}
\caption{Experimental results after finetuning in a new scenario. (a) NMSE in the new scenario. (b) NMSE in a previous scenario.}
\label{fig::results_finetune}
\vspace{-0.1cm}
\end{figure}

To evaluate the continual learning capabilities of different methods, we perform a few steps of finetuning using data from the new scenario Singapore BS 1. As shown in Fig. \ref{fig::results_finetune}(a), even after 60 finetuning steps, PCMixer struggles to achieve GCD's initial performance level in the new scenario. Furthermore, PCMixer exhibits catastrophic forgetting in previous scenarios with severe performance degradation, as shown in Fig. \ref{fig::results_finetune}(b). In contrast, GCD benefits from multi-scenario pre-training, enabling it to rapidly adapt to new scenarios while maintaining outstanding performance in previous ones.

\section{Conclusion}

In this work, we propose a channel acquisition framework that utilizes geometric information to provide scenario prompt. The coarse environmental map and BS position are used to extract geometric channel features, which are then aligned to the complete channel form using random placeholders, followed by fusion with partial estimates via a neural network. Comprehensive experimental validation demonstrates that our method achieves outstanding channel acquisition accuracy with sparse pilot resources and non-ideal geometric information, while exhibiting strong generalization capability in unseen scenarios and dynamic traffic environments.

\section*{Acknowledgment}

This work was supported in part by National Natural Science Foundation of China under Grants 62394292 and 624B2129, and in part by the Fundamental Research Funds for the Central Universities under Grant 226-2024-00069.

\printbibliography

\end{document}